\documentclass[aps,prx,showpacs,amsmath,amssymb,amsfonts,lengthcheck,twocolumn,longbibliography,superscriptaddress]{revtex4-2}
\usepackage[utf8]{inputenc}
\usepackage{soul}
\usepackage{natbib}
\usepackage{graphicx}
\usepackage{color}
\usepackage{amsfonts}
\usepackage{amsmath}
\usepackage[colorlinks=true,linkcolor=blue,urlcolor=blue,citecolor=blue,pdfusetitle]{hyperref}
\usepackage{txfonts}

\newcommand{\addition}[1]{#1}

\newcommand{\Ket}[1]{\left|#1\right>}
\newcommand{\Bra}[1]{\left<#1\right|}
\newcommand{\BraKet}[2]{\left<#1|#2\right>}

\newcommand{\trace}[1]{\mathrm{Tr}\left(#1\right)}
\newcommand{\traceb}[1]{\mathrm{Tr}\left[#1\right]}
\newcommand{\conm}[2]{\left[#1,#2\right]}
\newcommand{\aconm}[2]{\left\{#1,#2\right\}}
\newcommand{\mbf}[1]{\mathbf{#1}}
\newcommand{\tr}[1]{\mathrm{tr}\left\{#1\right\}}
\newcommand{\trp}[1]{\mathrm{tr}\left(#1\right)}
\newcommand{\ex}[1]{\exp{\left(#1\right)}}

\begin{document}

\title{Diverging quantum speed limits: A herald of classicality}
\author{Pablo M. Poggi}
\email{ppoggi@unm.edu}
\affiliation{Center for Quantum Information and Control, Department of Physics and Astronomy, University of New Mexico, Albuquerque, New Mexico 87131, USA}

\author{Steve Campbell}
\email{steve.campbell@ucd.ie}
\affiliation{School of Physics, University College Dublin, Belfield Dublin 4, Ireland}
\affiliation{\hbox{Centre for Quantum Engineering, Science, \& Technology, University College Dublin, Belfield, Dublin 4, Ireland}}

\author{Sebastian Deffner}
\email{deffner@umbc.edu}
\affiliation{Department of Physics, University of Maryland, Baltimore County, Baltimore, MD 21250, USA}
\affiliation{Instituto de F\'isica `Gleb Wataghin', Universidade Estadual de Campinas, 13083-859, Campinas, S\~{a}o Paulo, Brazil}

\begin{abstract}
When is the quantum speed limit (QSL) really \emph{quantum}? \addition{While vanishing QSL times often indicate emergent classical behavior,} it is still not entirely understood what precise aspects of classicality \addition{are at the origin of this dynamical feature}. Here, we show that vanishing QSL times (or, equivalently, diverging quantum speeds) can be traced back to reduced uncertainty in quantum observables \addition{and thus can be understood as a consequence of emerging classicality for these particular observables}. We illustrate this mechanism by developing a QSL formalism for continuous variable quantum systems undergoing general Gaussian dynamics. For these systems, we show that three typical scenarios leading to vanishing QSL times, namely large squeezing, small effective Planck’s constant, and large particle number, can be fundamentally connected to each other. \addition{In contrast,} by studying the dynamics of open quantum systems and mixed states, we show that \addition{the classicality that emerges due to incoherent mixing of states from the addition of classical noise typically increases the QSL time}.
\end{abstract}

\date{\today}
\maketitle
\section{Introduction}
What distinguishes the classical world from the underlying quantum domain? Arguably the most prominent answers to this question revolve around the existence of uncertainty relations. While these relations have been tested, understood, and verified for pairs of canonical observables, \addition{as there is no observable for time, the uncertainty relation for energy and time remains harder to interpret.} In its modern formulation the energy-time uncertainty is phrased {as a} quantum speed limit (QSL). In its original inception, the QSL {reads}~\cite{mand_tamm1945,fleming1973,bhatta1983} $\tau \geq \pi\hbar/2\Delta E\equiv \tau_{\rm QSL} $, where $\tau$ is the evolution time between orthogonal states, under a time-independent Hamiltonian, $H$, and $\Delta E^2\!=\! \Bra{\psi} H^2 \Ket{\psi} - \Bra{\psi} H \Ket{\psi}^2$.  QSLs have found widespread prominence in, e.g., quantum information theory~\cite{giovannetti2003,levitin2009}, while other formulations of QSLs provide fundamental and practical insight {into} the dynamics of complex systems~\cite{DeffnerReview, Frey2016, caneva2009, arenz2017, poggi2019, AlbertiPRX, AlbertiArXiv, delCampoPRL2021, PueblaPRR}. Formally, QSLs can be elegantly expressed in terms of the geometry of quantum evolution~\cite{anandan1990,pati1995,pires2016}, which in turn reveals a fundamental connection with the study of quantum parameter estimation~\cite{braunstein1994,giovannetti2006}. In this geometric setting, QSLs have been generalized and applied to {various} scenarios of interest, notably open quantum systems~\cite{taddei2013,deffner2013PRL,DelCampo2013} and quantum control~\cite{caneva2009,goerz2011,hegerfeldt2013,poggi2013,poggi2020}. 

Nevertheless, it is still debated what is really ``quantum'' about the QSL. Only recently, in two almost simultaneous works, Shanahan~\textit{et al}.~\cite{shanahan2018} and Okuyama \& Ohzeki~\cite{okuyama2018} showed that bounds resembling the QSL also exist for classical dynamics. The origin of such speed limits, quantum as well as classical, {rests} in the notion of \textit{distinguishability} of states. The speed limit is then a bound on the rate with which states become distinguishable from an previous configuration. While these results appear to put quantum and classical dynamics on equal footing, some differences are expected to persist. The natural question, thus, has to be if and how a diverging \emph{quantum} speed may be related to emergent classical behavior.

In this paper, we tackle this problem for a broad class of quantum systems, namely a collection of bosonic modes described by Gaussian Wigner functions under Gaussian-preserving dynamics~\cite{ferraro2005,adesso2014}. These systems provide an ideal testbed to study QSLs for both quantum and classical systems and have widespread applications in continuous variable (CV) quantum information~\cite{weedbrook2012}. In general, studying the QSL for CV systems is mathematically challenging due to their infinite-dimensional Hilbert spaces \cite{marian2021}. In contrast to previous work \cite{deffner2017,shanahan2018}, here we do not work with a {phase-space} representation, but rather develop a QSL theory for Gaussian dynamics directly, which permits to derive an expression for the QSL time in terms of finite dimensional matrices using symplectic operators.

Using this formalism, we discuss three limits in which the QSL time vanishes: {\it (i)} $\hbar\!\rightarrow\! 0$, where $\hbar$ is interpreted as a parameter of the state, {\it (ii)} $r\!\rightarrow\! \infty$, where $r$ denotes the squeezing in the state and {\it (iii)} $n\!\rightarrow \!\infty$ where $n$ is the number of modes. Thus, we establish that the emergent classicality linked to a vanishing QSL time can be associated to the reduced uncertainty in particular observables. For the special case of a single mode, we develop the theory further to show that, for each state, there exist Hamiltonians which maximize and minimize the QSL time. Finally, by applying our Gaussian QSL theory to general quantum evolution, we discuss the role of classical noise, mixed states and non-unitary evolution. We illustrate how these aspects, \addition{which are related to a transition to classical behavior due to incoherent mixing of states rather than reduced uncertainty of observables}, cannot decrease the QSL time. 

\section{Quantum speed limit for Gaussian dynamics.} We start by recalling the general formalism of geometric quantum speed limits. Consider a normalized distance between elements in the space of density operators given by $\Theta(\rho,\sigma)\!=\!2\arccos\sqrt{F(\rho,\sigma)}$, where $F(\rho,\sigma)$ is a fidelity function satisfying $0\!\leq\! F(\rho,\sigma) \!\leq\! 1$, and $F=1$ iff $\rho\!=\!\sigma$. Further, consider general quantum dynamics given by $\rho_t\!=\!\Lambda_t[\rho_0]$, where $\{\Lambda_t,\:t\geq0\}$ is a one-parameter family of completely-positive trace-preserving maps. \addition{The quantum speed $V_t$} is computed by expanding the fidelity between the state $\rho_t$ and the state at a subsequent time $\rho_{t+dt}$,
\begin{equation}
    F(\rho_t,\rho_{t+dt}) \equiv 1 - V_t^2\:dt^2  \Rightarrow d\Theta^2 = 4V_t^2\:dt^2.
    \label{ec:speed_fidelity}
\end{equation}
Note that $V_t$ measures how fast quantum states become distinguishable from each other. Generally, $V_t$ is a function of $\rho_t$, \addition{but may also show an explicit dependence in time}. Moreover, $V_t$ can be used to construct bounds on the evolution time in a variety of ways~\cite{poggi2013,mirkin2016,OConnorPRA}, typically based on \addition{the relation}
\begin{equation}
    \Theta(\rho_0,\rho_\tau) \leq 2 \int_0^\tau dt\, V_t.
    \label{ec:anandan_rel}
\end{equation}
Equation~\eqref{ec:anandan_rel} expresses the fact that the distance between $\rho_0$ and $\rho_\tau$ must be smaller or equal to the length of the path taken by $\rho_t$ for $t\!\in\!\left[0,\tau\right]$. 

For unitary dynamics and pure initial states the natural choice is the quantum fidelity $F\!=\!\lvert \BraKet{\psi_1}{\psi_2}\rvert^2$, for which $V_t\!=\!\Delta E_t/\hbar$ and Eq.~\eqref{ec:anandan_rel} is the Anandan-Aharonov relation ~\cite{anandan1990}. For time-independent Hamiltonians, the Mandelstam-Tamm bound can be easily inferred from Eq.~\eqref{ec:anandan_rel}, see also Ref.~\cite{Deffner2013JPA}. 

In the present analysis, we focus on the speed for Gaussian states and define the QSL time simply as its inverse, i.e., $\tau_Q\!\equiv\! V^{-1}$. \addition{The time-dependence of $V$ has been dropped, since we can take $V$ to be the speed at the initial time $t=0$. It is straight forward to show (see Appendix \ref{app:definitions_qsl}) that the other usual definitions of a QSL time are analogous to $\tau_Q$ in the asymptotic limit $\tau_Q \rightarrow0$}.  Generic bosonic systems have a clear-cut classical limit, which makes them ideal to study the quantum-to-classical transition. Gaussian-preserving dynamics \addition{in systems of $n$ modes} can be efficiently described using finite-dimensional operators corresponding to the symplectic group $\mathrm{Sp}(2n)$~\cite{hall2015,adesso2014}. These systems can be characterized by a vector of quadrature operators $\mathbf{\hat{z}}\!=\!(\hat{q}_1,\hat{p}_1,\ldots,\hat{q}_n,\hat{p}_n)$ with commutation relations \footnote{Since we are interested in studying the role of $\hbar$ in the QSL, we have defined the quadrature operators to be independent of $\hbar$. For the typical harmonic oscillator Hamiltonian, $H_{\mathrm{HO}}\!=\!P^2/2m+m\omega^2/2\,Q^2$, the definition used in the main text corresponds to taking $q\!=\!\sqrt{m\omega}\,Q$ and $p=P/\sqrt{m\omega}$ such that $\conm{q}{p}\!=\!\conm{Q}{P}=i\hbar$} \addition{\begin{equation}
[\hat{z}_k,\hat{z}_l]\! =\! i \hbar\, \Omega_{kl},\quad\text{where}\quad \Omega=\bigoplus_{j=1}^n \left(\begin{array}{c c}
    0 & 1 \\
    -1 & 0 \end{array}\right)\,.
\end{equation}}
    
A state, $\rho_G$, is Gaussian if its Wigner distribution is a Gaussian function in the quadrature variables. These states can be fully described by a $2n$-dimensional real vector of expectation values $\mathbf{u}\!=\!\trp{\hat{\rho}_G\, \mathbf{\hat{z}}}$ and a real, symmetric $2n\!\times\!2n$ covariance matrix $\Sigma \!=\! \trp{\hat{\rho}_G\, \{\delta \mathbf{\hat{z}},\delta \mathbf{\hat{z}}^T\}}$, \addition{where $\delta \mathbf{\hat{z}}\equiv \mathbf{\hat{z}}-\mathbf{u}$}, which is such that $\Sigma + i\hbar\,\Omega\! \geq\! 0$. {Thus, the} purity of $\rho_G$ is {simply} given by 
\addition{\begin{equation}
\tr{\rho_G^2}=\sqrt{\hbar/\det(\Sigma)}\,.
\end{equation}}

For mathematical convenience, we now choose the metric, $F$, to be the fidelity introduced in~\cite{Wang2008,Sun2015}
\begin{equation}
    F(\rho,\sigma)=\frac{\tr{\rho\, \sigma}}{\sqrt{\tr{\rho^2}\tr{\sigma^2}}}\,.
    \label{ec:alt_fidelity}
\end{equation}
Note that for purity-preserving dynamics, Eq. (\ref{ec:alt_fidelity}) reduces to the relative purity~\cite{DelCampo2013} and the corresponding distance, $\Theta$, to the one studied in~\cite{campaioli2018}. For a CV system, $\rho$ is represented by an infinite-dimensional matrix, however, Gaussian states can be described by finite-dimensional objects $\rho\! \rightarrow\! (\Sigma,\mathbf{u})$. \addition{In terms of these, the fidelity of Eq.~(\ref{ec:alt_fidelity}) can be expressed as~\cite{marian2012,link2015} }
\begin{equation}
\begin{split}
    F\left(\rho_1,\rho_2\right) =& \frac{\left(\det(\Sigma_1)\right)^{1/4}\left(\det(\Sigma_2)\right)^{1/4}}{\left(\det(\frac{1}{2}\left(\Sigma_1+\Sigma_2)\right)\right)^{1/2}} \\ 
    & \times \mathrm{exp}\left(-\delta \mbf{u}^T (\Sigma_1+\Sigma_2)^{-1}\delta \mbf{u}\right),
    \label{ec:fide_gaussians}
\end{split}
\end{equation}
where $\delta \mbf{u} \!\equiv\! \mbf{u}_2 - \mbf{u}_1$. \addition{Consider now a transformation $\rho\rightarrow \rho+d\rho$.} If the evolution preserves the Gaussian character of the state, we can expand Eq.~\eqref{ec:fide_gaussians} to second order using $\Sigma\!\rightarrow\!\Sigma+d\Sigma$ and $\mbf{u}\!\rightarrow\! \mbf{u}+d\mbf{u}$. This procedure, \addition{detailed in Appendix \ref{app:derivation}}, yields the expression
\begin{equation}
\begin{split}
    F(\rho,\rho+d\rho) &= 1 - \frac{1}{16}\,\tr{\left(\Sigma^{-1}d\Sigma\right)^2}\\
    &\qquad- \frac{1}{2}\,d\mbf{u}^T \Sigma^{-1} d\mbf{u},
\end{split}
    \label{ec:dif_fide_gaussian}
\end{equation}

\noindent \addition{which we will use to obtain explicit expressions for the quantum speed $V$ for different cases of interest}. 

We will first focus on unitary evolutions generated by quadratic Hamiltonians of the form $H\!=\!\mbf{z}^T G\mbf{z}/2$, with $G\!\in\! \mathbb{R}^{2n\times2n}$ and symmetric. In this case the quadrature operators evolve according to  $\mbf{z}(t)\!=\!S(t)\mbf{z}(0)$, with $S(t)$ a symplectic matrix (i.e., such that $S\Omega S^T\!=\!\Omega$) obeying $\dot{S}\!=\!\Omega G S(t)$, which in turn implies that initial Gaussian states remain Gaussian at all times. \addition{The unitary evolution of the covariance matrix and the displacement vector of a Gaussian state is given by 
\begin{equation}
\addition{d\Sigma=(\Omega G\Sigma-\Sigma G \Omega)\,dt\quad\text{and}\quad d\mbf{u}=\Omega G\mbf{u}\,dt}\,.
\end{equation} }
Inserting this into Eq.~\eqref{ec:dif_fide_gaussian} and using Eq.~\eqref{ec:speed_fidelity}, we can evaluate the quantum speed for unitary evolution of general (multimode, mixed) Gaussian states,
\begin{equation}
\begin{split}
V^2_{\mathrm{U}} &= \frac{1}{8}\,\left(-\tr{G\Omega\Sigma^{-1}\Omega G\Sigma }+\tr{(\Omega G)^2}\right)\\
&\qquad-\frac{1}{2}\,\mbf{u}^T G \Omega\Sigma^{-1}\Omega G \mbf{u}.
\end{split}
\label{ec:speed_unitary}
\end{equation}
This expression can be further simplified if $\rho\!\rightarrow\! \{\Sigma, \mbf{u}\}$ is pure. In this case, using Williamson's theorem it is straightforward to show that $-\Omega \Sigma^{-1} \Omega =\Sigma/\hbar^2$ \addition{(see Appendix \ref{app:speed_pure}). This leads to the following general expression for the quantum speed for a generic pure Gaussian state undergoing Gaussian-preserving dynamics} 
\begin{equation}
\begin{split}
   V^2_{\mathrm{U}} &= \frac{1}{8\hbar^2}\,\left(\tr{(G\Sigma)^2}+\hbar^2\,\tr{(\Omega G)^2}\right)\\
   &\qquad+\frac{1}{2\hbar^2}\,\mbf{u}^T G\Sigma G \mbf{u}.
\end{split}
  \label{ec:speed_unipure}
\end{equation}
\addition{As expected from the Anandan-Aharonov relation, the expression in Eq. (\ref{ec:speed_unitary}) coincides with the energy variance $\Delta E^2/\hbar^2$, a fact we prove by direct calculation in Appendix \ref{app:speed_pure}}. In the following, we turn our attention to analyzing the different limits in which the quantum speed diverges, leading to a vanishing QSL time.

\begin{figure}
 \centering{\includegraphics[width=0.8\linewidth]{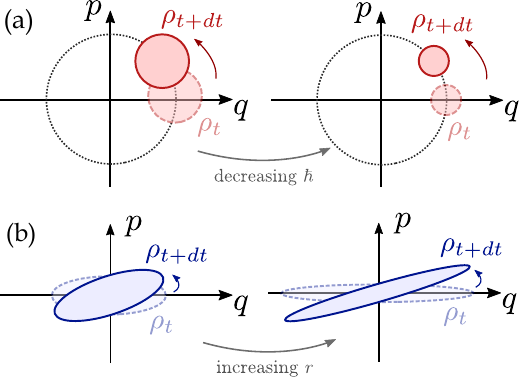}}
\caption{Pure Gaussian states undergoing an infinitesimal evolution in phase space quickly become distinguishable if (a) $\hbar$, taken as a parameter of the state, is reduced, and (b) the squeezing parameter, $r$, is increased.}
\label{fig:vanishing}
\end{figure}

\section{Diverging quantum speed limits}

\subsection{Limit of small Planck's constant} 
We begin by analyzing the role of $\hbar$ in the QSL time. It is instructive to evaluate Eq.~\eqref{ec:speed_unipure} for a generic multimode state evolving on a uniform Harmonic oscillator, where $G\!=\!\omega\, \mathbb{I}_{2n\times 2n}$. The covariance matrix of an arbitrary Gaussian pure state can be written as $\Sigma\!=\! \hbar\, O D O^T$, where $O$ is an orthogonal matrix and $D$ is a diagonal positive matrix of the form
\addition{
\begin{equation}
    D\!=\!\bigoplus\limits_{k=1}^{n} \left(\begin{array}{c c}
    x_k & 0 \\
    0 & 1/x_k \end{array}\right).
\end{equation}
}
The elements of $D$ describe the \addition{magnitude of the squeezing} of the state and can be parametrized as $x_k\!=\ex{r_k}$, with $r_k\!\geq\! 0$.  The resulting expression for the speed reads
\begin{equation}
    V^2_{\mathrm{U}}=\frac{\omega^2}{8}\,\left(\tr{D^2}-2n\right) + \frac{\omega^2}{2\hbar}\, \mbf{v}^T D \mbf{v} 
    \label{ec:speed_harmonic}
\end{equation}
where $\mbf{v}\!=\!O^T \mbf{u}$. Equation~\eqref{ec:speed_harmonic} exhibits two distinct contributions. The first term corresponds to the speed originating in the squeezing of the state (and it vanishes in its absence, i.e., when $x_k\!=\!1\:\forall k$); the second term indicates the displacement of the state. 

Observe that the speed, $V^2_{\mathrm{U}}$, diverges as $\hbar\!\rightarrow\! 0$ \cite{BolonekLason2021}. This happens only for displaced states, since the first term is independent of $\hbar$. This behavior can be understood in terms of the evolved state becoming more distinguishable from the initial one as $\hbar$ is reduced. This is depicted for the single mode in Fig.~\ref{fig:vanishing}(a). The necessary counterpart of $\hbar\!\rightarrow\! 0$ is that the state becomes more classical in the sense that the uncertainty in {\it all} quadratures is reduced. Thus, there is a straightforward connection between a vanishing QSL time and a reduced uncertainty associated with the state of the system. \addition{This result makes explicit the fact that the role of $\hbar$ in the QSL is precisely to set the minimum uncertainty, which limits the rate of change of the distinguishability. Since uncertainty can always be introduced in classical systems, a similar mechanism can  be understood to lead to a speed limit for those systems~\cite{shanahan2018,okuyama2018}.}

\subsection{Limit of large squeezing} 
We now turn to the second limit. For simplicity, we set without loss of generality $\mbf{u}\!=\!0$, \addition{and therefore restrict to considering states which are centered at the origin in phase space}. Using the same example as above, we have that 
\addition{\begin{equation}
\tr{D^2}\!=\!2\sum_{k=1}^n \cosh(2r_k)\!=\!4\sum_{k=1}^n \sinh^2(r_k)+2n
\end{equation}}
which leads to
\begin{equation}
V^2_{\mathrm{U}}\!=\!\frac{\omega^2}{2}\,\sum_{k=1}^n \sinh^2(r_k) 
\end{equation}
revealing that, as the squeezing of the state increases, the quantum speed diverges. This phenomenon can again be rationalized from the fact that large squeezing allows for a faster increase in distinguishability, as schematically depicted in Fig.~\ref{fig:vanishing}(b). Here we observe that a diverging speed is again associated with reduced uncertainty, in this case corresponding to the variance of the squeezed quadrature operator(s) of the state.

\addition{An important observation about the role of states and  generators in the QSL follows from this example. Since the quantum speed is the rate at which the state of the system becomes distinguishable from its previous configuration under a given evolution, a vanishing QSL time can be achieved trivially if the generator (Hamiltonian in the unitary case) itself is unbounded (i.e. $\omega\rightarrow \infty$ in Eq. (\ref{ec:speed_harmonic})). What we have shown here is that a vanishing QSL time with a bounded generator is also possible, even in the case of a system of single mode ($n\!=\!1$), provided the state is a highly squeezed state with $r\!\rightarrow\! \infty$.} This is a feature of CV systems which is absent in the finite-dimensional case where the quantum speed is strictly upper bounded by the norm of the Hamiltonian~\cite{brody2015,poggi2020} and thus vanishing QSL times are prohibited (for fixed $\hbar$). 

\begin{figure} [h!]
\centering
\includegraphics[width=\linewidth]{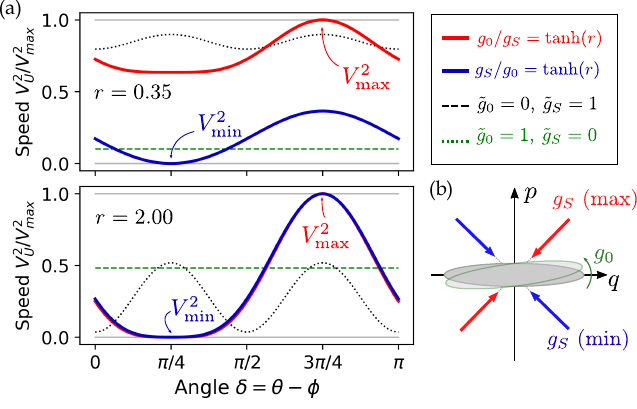}
    \caption{Plots in (a) show $V^2_\mathrm{U}$, normalized to its maximum value, as a function of $\delta=\theta-\phi$, where $\theta$ is the angle of the squeezed quadrature and $\phi$ is the angle characterizing the squeezing Hamiltonian, cf. Eq.~(\ref{ec:hami_single}). For a given degree of squeezing (top plot $r=0.35$, bottom plot $r=2.00$), the maximum speed is achieved at $\delta=3\pi/4$, when $g_0/g_S=\tanh(r)$, while the minimum is reached when $g_S/g_0=\tanh(r)$. These conditions give $g_S\simeq g_0$ for large $r$, as can be seen by the curves in the bottom plot. Other choices of $g_0,g_S$ are shown for comparison. (b) Schematic showing the action of the optimal Hamiltonians which maximize (red) and minimize (blue) the speed of a state squeezed along the $p$ quadrature. The arrows indicate the direction of squeezing.}
    \label{fig:singlemode}
\end{figure}

We investigate this behavior further by fully characterizing the quantum speed for generic, quadratic single mode Hamiltonians. The complete derivation is relegated to \addition{Appendix \ref{app:singlemode}}. For $n\!=\!1$, we can write a general mixed state as $\Sigma\!=\!\hbar c\, ODO^T$, where $O$ is a rotation matrix by an angle $\theta$, and $c\!>\!0$. The generator, $G$, becomes
\begin{equation}
    G = g_0 G_0+ g_S\left(\sin(2\phi)\,G_1+\cos(2\phi)\,G_2\right)
    \label{ec:hami_single}
\end{equation}
where $g_0,\:g_S \!\in\! \mathbb{R}$ are the weights corresponding to the number-preserving and number-non-preserving parts of the generator. $G_0$, $G_1$ and $G_2$ are the $2\!\times\!2$ matrix representations of the single-mode Gaussian-preserving (quadratic) Hamiltonians $q^2+p^2$, $q^2-p^2$ and $qp+pq$, respectively. \addition{The angle $\phi$ is introduced to parametrize the relative contribution of each of the squeezing generators $G_1$ and $G_2$}. In this case, $V_U$ can be evaluated exactly, yielding  
\begin{equation}
\begin{split}
V_\mathrm{U}^2&=\frac{1}{2}\,\left[\left(g_0\, \sinh(r)-g_S\,\sin(2\delta)\cosh(r)\right)^2\right.\\
&\qquad \left.+g_S^2\, \cos^2(2\delta) \right],
\end{split}
\label{ec:vel_singlemode}
\end{equation}
where we have introduced $\delta\!=\!\theta-\phi$. 

The speed, $V^2_{\mathrm{U}}$, is plotted in Fig.~\ref{fig:singlemode} for various combinations of parameters. The maximum speed, $V^2_{\mathrm{max}}$, occurs when $\delta \!=\! 3\pi/4$. Introducing an overall energy scale $g$ such that $g_0 \!=\! g\tilde{g_0}$ and $g_S\!=\!g \tilde{g_S}$, we have that $V^2_{\mathrm{max}}\!=\!g^2\,\cosh(2r)/2$, which grows as $\ex{2r}$ for large $r$ and for $g_0/g_S\!=\!\tanh(r)$. For low squeezing, this amounts to setting $g_S\!\gg\! g_0$, while for high squeezing, it is achieved by $g_S\!\simeq\! g_0$. Consequently, for a single mode system there always exists a Hamiltonian for which the QSL time vanishes optimally as $\ex{-2r}$ when the squeezing is large. It can also be shown that for any degree of squeezing, an ``opposite'', minimum-speed Hamiltonian exists, for which $V_\mathrm{U}$ is either zero or independent of $r$, see \addition{Appendix~\ref{app:singlemode}}.

\subsection{Limit of large system size} 
Finally, we analyze the large $n$ limit. Taking $\mbf{v}\!=\!(\langle\tilde{q}_1\rangle,\langle\tilde{p}_1\rangle,\ldots,\langle\tilde{q}_n\rangle,\langle\tilde{p}_n\rangle)$, Eq.~\eqref{ec:speed_harmonic} becomes 
\begin{equation}
\begin{split}
    &V^2_{\mathrm{U}}=\frac{\omega^2}{2}\left[ n\left(\frac{1}{n}\sum\limits_{k=1}^n\sinh^2(r_k)\right) \right. \\
    &\left. + \frac{n}{\hbar} \left( \frac{1}{n}\sum\limits_{k=1}^n \ex{r_k} \langle\tilde{q}_k\rangle^2 + \frac{1}{n}\sum\limits_{k=1}^n \ex{-r_k} \langle\tilde{p}_k\rangle^2\right) \right].
    \label{ec:speed_largen}
\end{split}
\end{equation}
Assuming that the squeezing parameters $\{r_k\}$ and the displacements $\{\langle\tilde{q}_k\rangle,\langle\tilde{p}_k\rangle\}$ are independent of $n$, the quantities in the parenthesis of Eq.~\eqref{ec:speed_largen} remain intensive as $n\!\rightarrow\! \infty$ and thus we find $V^2_\mathrm{U}$ diverges linearly with $n$. This behavior of $V_\mathrm{U}$ bears close resemblance to the origin of the orthogonality catastrophe studied in Ref.~\cite{fogarty2020}.

Interestingly, the role of $n$ in the vanishing QSL time can be recast in terms of the two limits studied above. We begin with the first term in Eq.~\eqref{ec:speed_largen}. For fixed $r_k$'s, the resulting multimode speed can be emulated by a single mode system where the squeezing parameter $r$ obeys $\sinh^2(r)\!=\!\sum_k \sinh^2(r_k)$. For small $r_k\!=\!r^{(n)}$, we have that $r\!\simeq\! \sqrt{n}\: r^{(n)}$, and thus, in absence of displacements, we find that the large $n$ limit with finite squeezing is equivalent to the large squeezing limit of a single mode evolution. We can include displacements in the analysis by now considering emulating the second term in Eq.~\eqref{ec:speed_largen} with a single mode system. The components of the required displacement vector read
\begin{equation}
\begin{split}
    \langle \tilde{q}\rangle &=  \frac{\ex{-r}}{n}\sum_{k=1}^n \ex{r_k} \langle \tilde{q}_k\rangle^2\,\\
    \langle \tilde{p}\rangle &=  \frac{\ex{r}}{n}\sum_{k=1}^n \ex{-r_k} \langle \tilde{p}_k\rangle^2\,.
\end{split}
\end{equation}
In the absence of squeezing, $r_k\!=\!r=0$, the displacement vector length does not scale with $n$. Thus, the limit $n\!\rightarrow\! \infty$ is equivalent to letting $\hbar_\mathrm{eff}\!\equiv\! \hbar/n \!\rightarrow\! 0$, and thus it reduces to the first case considered above. In the presence of squeezing, the length of the displacement vector increases with $n$ since $\ex{r}\!\sim\! \ex{\sqrt{n}}$. Thus, we can further normalize $\mbf{v}\!\rightarrow\! \mbf{v}/\cosh(r)$ such that $||\mbf{v}||^2 \!\sim\! \ex{r}/\cosh(r) \!\simeq\! 1$ as $n\!\rightarrow\! \infty$. Therefore, $\hbar_{\mathrm{eff}} \!\equiv\! \hbar / (n\cosh{r}) \!\sim\! \hbar/n^2$, which vanishes for $n\gg 1$.

\begin{figure}
\centering
       \includegraphics[width=\linewidth]{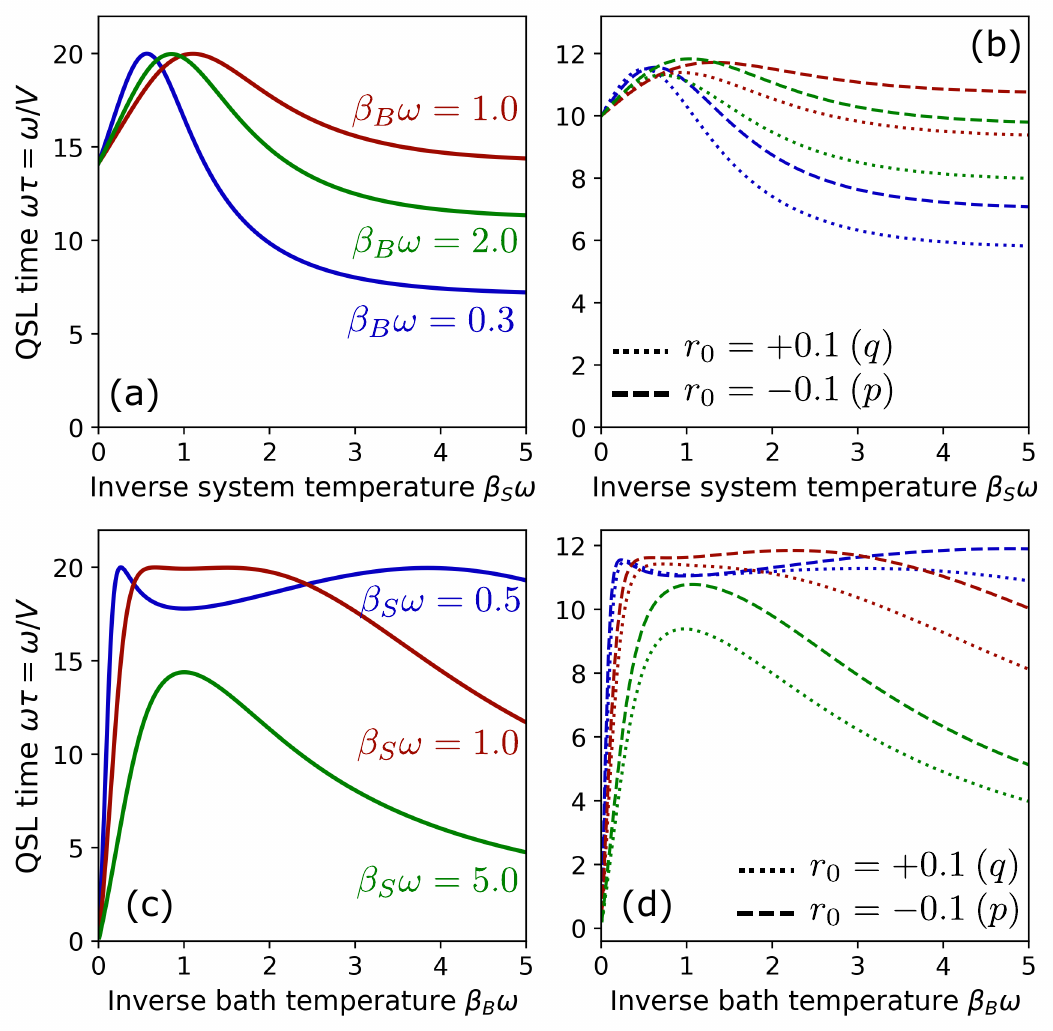}
 \caption{Plots (a) and (b) depict the QSL time $\tau = V^{-1}_\mathrm{cov}$ as a function of the inverse effective \textit{system} temperature $\beta_S$ for different values of the \textit{bath} temperature $\beta_B$. (a) $r=0$ (vacuum) and (b) $r=\pm 0.1$ (squeezing along $q$ or $p$). In these cases the QSL time is always nonzero. Plots (c) and (d) show the converse cases:  $\tau = V^{-1}_\mathrm{cov}$ as a function of $\beta_B$, for fixed values of system temperature. In all cases $\gamma=1$. }
    \label{fig:fig_qbm}
\end{figure}

\section{Mixed states and non-unitary evolution} 
So far we have analyzed the quantum speed of evolution for pure Gaussian states, and we have shown a relation between the diverging speed and a particular aspect of the classicality of the state, i.e. the uncertainty of an observable (or set of observables) vanishing. A seemingly separate notion of classicality is given by considering mixed states and purity-non-preserving evolution. Mixed states are classical mixtures of pure states and the addition of classical noise is expected to reduce distinguishability~\cite{shanahan2018,okuyama2018}. 

To elucidate the matter, we now generalize our QSL theory of Gaussian states for general open quantum dynamics. Equation~\eqref{ec:dif_fide_gaussian} can be applied to study any dynamics that preserves the Gaussian character of the state, such as general open diffusive dynamics~\cite{genoni2016}. Here we focus on the single mode case ($n\!=\!1$), which allows us to treat the most general Gaussian-preserving evolution in an exact way. The equations of motion can be written as (see Appendix \ref{app:open_quantum} for further details)
\begin{equation}
\begin{split}
    \dot{\Sigma} &= \Omega G\Sigma - \Sigma G\Omega - g\,\left(\Sigma-M\right)\,,\\
    \dot{\mbf{u}}&=\left(\Omega G-g/2\,\mathbb{I}\right)\mbf{u}\,, \label{ec:open_eqs}
\end{split}
\end{equation}
where $g\!\in\! \mathbb{R}$ and $M\!\in\! \mathbb{R}^{2\times 2}$. As expected from Eq.~\eqref{ec:dif_fide_gaussian}, the quantum speed has the form 
\addition{\begin{equation}
    V^2_{\rm open} \!=\! V^2_{\rm cov} + V^2_{\rm mean}\,,
\end{equation}}
with the first term stemming solely from the covariance matrix, and the second one from the evolution of the mean values. Focusing on the former, we obtain that $V^2_{\rm cov}\!=\!V^2_{\rm U} + \chi_{\rm NU}$ where 
\begin{equation}
    V^2_{\rm U} = \frac{1}{8}\,\left(\frac{1}{\eta^2}\tr{(G\Sigma)^2}+\tr{(\Omega G)^2}\right),
    \label{ec:speed_open_uni}
\end{equation}
is the contribution from unitary dynamics, i.e., the generalization of Eq.~\eqref{ec:speed_unitary} for single-mode mixed states. Further, we introduced $\eta\!=\! \sqrt{\det(\Sigma)}$, and
\begin{equation}
\begin{split}
    \chi_{\rm NU} =& \frac{g^2}{8}\,\left(1-\tr{\Sigma^{-1}M}+\tr{\left(\Sigma^{-1} M\right)^2}/2\right) \\
    &+ \frac{g}{8\eta^2}\,\tr{\Sigma\left(GM\Omega-\Omega MG\right)},
\end{split}
\label{ec:speed_open}
\end{equation}
is the contribution from the nonunitary part of the dynamics. For any choice of evolution given by $G$, $M$ and $g$, we can evaluate the speed for a squeezed thermal state $\Sigma \!=\! \eta(\beta_s)\, O D O^T$, where we have introduced the effective inverse temperature of the state $\beta_s$ via the usual parametrization  $\eta(\beta_s)\!=\!2\bar{n}_s+1\!\equiv\! \coth\left(\beta_s\omega/2\right)$. From Eq.~\eqref{ec:speed_open_uni} it becomes evident that $V^2_{\rm U}$ is independent of $\beta_s$. The non-unitary contribution becomes
\begin{equation}
\begin{split}
    &\chi_{\rm NU} = \frac{g^2}{8}\,\left[1-\frac{1}{\eta}\,\tr{OD^{-1}O^TM} \right. \\
   &\quad\left. + \frac{1}{g\eta}\,\tr{ODO^T\left(GM\Omega-\Omega MG\right)} \right]+\mathcal{O}(\eta^{-2}).
\end{split}
\end{equation}
At large temperatures (small $\beta_s$), where classical noise dominates, we have  $\eta\!\sim\! (\beta_s\omega)^{-1}$ and thus only the first term in $\chi_{\rm NU}$ survives. In this limit, the speed has an asymptotic, finite value $V^2_{\rm cov} \!=\! V^2_U + g^2/8$, thus confirming that increasing classical noise always yields a nonzero speed limit time.

\paragraph*{Quantum Brownian motion.} As a last point, we explore the effects of the \textit{bath} temperature on the quantum speed of evolution and compare its role with respect to the system's effective temperature. To this end, we focus on quantum Brownian motion (QBM), which describes the dynamics of a single harmonic oscillator interacting with a bosonic bath \cite{hu1992,schlosshauer}. The master equation reads
\begin{equation}
    \dot{\rho} = -i\conm{H_0}{\rho}-\Delta\conm{q}{\conm{q}{\rho}}+\Pi \conm{q}{\conm{p}{\rho}}-i\gamma \conm{q}{\aconm{p}{\rho}}
\end{equation}
where $H_0 \!=\!\omega\,(q^2+p^2)/2$. At high temperatures, the diffusion coefficients $\Delta$ and $\Pi$ can be written in terms of the damping rate $\gamma$ as $\Delta\!=\!{\gamma}/{\beta_B}+{12\gamma}(\omega^2-\gamma^2)/{\beta_B}$ and $\Pi\!=\!-\gamma\beta_B/12$, where $\beta_B$ is now the inverse temperature of the bosonic bath~\cite{deffner2013,deffner2017}. The dynamics of the QBM is Gaussian-preserving~\cite{vasile2009,torre2018} and thus can be cast in the form of Eq.~\eqref{ec:open_eqs} where $G\!=\!\omega \mathbb{I}$, $g\!=\!2\gamma$ and 
\begin{equation}
 M \! =\! \frac{1}{\gamma}\left(\begin{array}{c c}
    \Delta & -\Pi/2 \\
    -\Pi/2 & 0 \end{array}\right).
\end{equation}

To analyze the role of $\beta_B$ and $\beta_S$, we plot in Fig.~\ref{fig:fig_qbm}  the QSL time $\tau\! =\! V_{\rm cov}^{-1}$. In Fig.~\ref{fig:fig_qbm}(a), $\tau$ is plotted for fixed bath temperature $\beta_B$ as a function of $\beta_S$ and for no squeezing, while in (b) the same is shown for the case of squeezing along $q$-quadrature, or squeezing along $p$-quadrature (dashed and dotted lines, respectively). As expected, we observe that the QSL time remains nonzero in all cases. Furthermore, $\tau$ reaches a bath-independent value at high temperature ($\beta_S\!\rightarrow\! 0$), while it decays to a bath-dependent regime at low temperatures. In Fig.~\ref{fig:fig_qbm}(c) and (d) the QSL time is plotted as a function of the inverse \textit{bath} temperature, for fixed values of $\beta_B$, and the same squeezing regimes as before. The resulting behavior is notably different, since $\tau$ vanishes at high bath temperature for all cases. The shape of these curves can be understood by analyzing the expression of the speed in the case with no squeezing $r\!=\!0$ (which leads to $V^2_{\rm U}\!=\!0$). There, we obtain 
\begin{equation}
    V^2_{\rm QBM} = \frac{\gamma^2}{2}\left(1-x(\beta_S,\beta_B)+x(\beta_S,\beta_B)^2/2\right) + \mathcal{O}(\beta_B)
    \label{ec:qbm_high}
\end{equation}
where $x(\beta_s,\beta_B)\!=\!\beta_B^{-1} \tanh\left(\beta_S \omega/2\right)$. For fixed $\beta_B$, the speed remains bounded for all $\beta_S$, and reaches a \addition{bath-independent} value of $\gamma^2/4$ at large temperatures $\beta_S\ll 1$. This illustrates the behavior discussed above, where the system's effective temperature, related to the mixed nature of the state, cannot increase the speed arbitrarily and thus will not lead to vanishing QSL times. The role of the \textit{bath} temperature is markedly different since, for fixed $\beta_S$, we get that $V^2_{\rm QBM}$ grows unboundedly as $T_B^2$. \addition{Note, however, that $T_B$ is a property of the generator of the non-unitary evolution, and throughout this work we have focused on vanishing QSL times for \textit{bounded} generators}.

\addition{Finally, other interesting features of the curves in Fig. \ref{fig:fig_qbm} can be deduced from Eq. (\ref{ec:qbm_high}). First, notice that the curves are not monotonic, and in particular the ones in (a) and (b) display a peak for a given combination of $\beta_B$ and $\beta_S$. This behavior is captured by Eq. (\ref{ec:qbm_high}) which predicts a minimum of the speed (maximum of QSL time) at $x(\beta_S,\beta_B)=1$. This condition corresponds to the case where the temperature of the system is roughly equal to the temperature of the bath (assumed to be large here). Under these conditions, the QBM dynamics has a steady state which is roughly equilibrated with the bath, and the closer the system state is to this equilibrium state, the smaller the speed. For arbitrary $\beta_B$, this same mechanism explains the emergence of the peaks, albeit the relation between the equilibrium temperature and the bath temperature is more intrincate. On the other hand, Eq. (\ref{ec:qbm_high}) also predicts the plateau behavior seen at large $\beta_S$ in Figs. \ref{fig:fig_qbm} (a) and (b). For $x\rightarrow \infty$ (and in the regime of small $\beta_B$, where this expression is valid), we obtain a plateau value of $V^2_{\mathrm{QBM}}\simeq 1/(2\beta_B^2)$, which becomes larger as the bath temperature increases (and thus the state is further away from the equilibrium configuration).}

\section{Concluding remarks}
We have shown that vanishing QSL times in continuous variable systems can be traced back to an underlying property: the asymptotically vanishing uncertainty of a set of particular observables which depend on the state and the dynamics. \addition{This result shows that a very particular notion of classicality, strictly related to this vanishing uncertainty, is responsible for the absence of a QSL. This property can emerge for systems as simple as a single bosonic mode in a highly squeezed state, but is absent in quantum systems with finite-dimensional Hilbert spaces. By studying the behavior of the QSL in open quantum systems, we have explored the behavior of other aspects of classicality on the QSL, and showed that, in contrast, the addition of classical noise, be it from considering mixed states, or from dissipative dynamics, will not lead to vanishing QSL times.} To derive these results, we have developed a QSL framework for continuous variable systems undergoing Gaussian-preserving dynamics. We expect this framework to have broader applications, particularly in the study of quantum control of CV systems and non-Markovianity~\cite{wu2008,jahromi2020}. 

\acknowledgments{This material is based upon work supported by the U.S. Department of Energy, Office of Science, National Quantum Information Science Research Centers, Quantum Systems Accelerator. S.C. gratefully acknowledges the Science Foundation Ireland Starting Investigator Research Grant ``SpeedDemon" (No. 18/SIRG/5508) for financial support. S.D. acknowledges support from the U.S. National Science Foundation under Grant No. DMR-2010127.}

\bibliography{references}

\pagebreak
\widetext
\appendix
\section{Different definitions of QSL time} \label{app:definitions_qsl}

In the main text we define the QSL time as $\tau_Q=V^{-1}$, however a more common definition is given by  
\begin{equation}
    \tau_1(t) = \frac{\theta(\rho_0,\rho_{t})}{1/t\,\int_0^{t} dt'\, V(t')}.
\end{equation}
An alternative definition is given by $\tau_2(t)$ which is implicitly defined by the equation \cite{poggi2013,mirkin2016}
\begin{equation}
    \theta(\rho_0,\rho_t)=\int_0^{\tau_2(t)}dt'\,V(t').
\end{equation}
In both cases, Eq. (\ref{ec:anandan_rel}) ensures that $t\geq \tau_i$. For short $t,\tau_i$, one can approximate $V(t)$ as constant (i.e, evaluated at $\rho_0$, and thus from the expressions above one gets
\begin{equation}
    V(t)\tau_i=\theta(\rho_0,\rho_t) \rightarrow \tau_i \propto V^{-1},
\end{equation}
thus being proportional to $\tau_Q$ as intended.

\section{Derivation of the quantum speed} \label{app:derivation}
Here we derive Eq.  (\ref{ec:dif_fide_gaussian}) in the main text, which gives the expression for the quantum speed $V_t$ associated with the fidelity $F(\rho_1,\rho_2)$. First, take the Gaussian states $\rho_1\rightarrow (\Sigma,\mbf{u})$ and $\rho_2\rightarrow (\Sigma+d\Sigma,\mbf{u}+d\mbf{u})$. Using Eq. (\ref{ec:fide_gaussians}), we get
\begin{equation}
    F\left(\rho_1,\rho_2\right)=\frac{\det\left(\mathbb{I}+\Sigma^{-1}d\Sigma\right)^{1/4}}{\det\left(\mathbb{I}+\frac{1}{2}\Sigma^{-1}d\Sigma\right)^{1/2}}\,\ex{-d\mbf{u}^T(2\Sigma+d\Sigma)^{-1} d\mbf{u}}
    \label{ec:app_fide1}
\end{equation}
where we exploit the fact that $\Sigma$ is invertible and properties of the determinant. Consider the first factor in the equation above, which is the ratio of two expressions of the form $\det\left(\mathbb{I}+\alpha B\right)^\gamma$. Any matrix $A$ obeys $\det A=\ex{\trace{\log\left(A\right)}}$, and so one can expand the determinant to obtain
\begin{eqnarray}
\det\left(\mathbb{I}+\alpha B\right)^\gamma&=&\ex{\gamma \tr{\alpha B -\frac{1}{2}\alpha^2 B^2} + \mathcal{O}(B^3)} \\
&=& 1 + \alpha \gamma \tr{B} - \frac{\gamma \alpha^2}{2}\tr{B^2}+\frac{\alpha^2\gamma^2}{2}\tr{B}^2 + \mathcal{O}(B^3)
\label{ec:app_expansion}
\end{eqnarray}
Using the general expression Eq. (\ref{ec:app_expansion}) we can expand the first factor in Eq. (\ref{ec:app_fide1}) in a straightforward way. The result reads

\begin{equation}
    \frac{\det\left(\mathbb{I}+\Sigma^{-1}d\Sigma\right)^{1/4}}{\det\left(\mathbb{I}+\frac{1}{2}\Sigma^{-1}d\Sigma\right)^{1/2}} \simeq 1-\frac{1}{16}\tr{B^2},\ \mathrm{where}\ B=\Sigma^{-1}d\Sigma.
    \label{ec:app_factor1}
\end{equation}

The second factor in Eq. (\ref{ec:app_fide1}) is of the form $\ex{-d\mathrm{u}^T L^{-1} d\mathrm{u}}$, where $L=2\Sigma + d\Sigma$. Expanding the inverse, one finds that up to terms that are quadratic in $(d\mbf{u},d\Sigma)$, the leading term corresponds to $L^{-1}\simeq \Sigma^{-1}/2$. The resulting expansion reads 
\begin{equation}
    \ex{-d\mbf{u}^T(2\Sigma+d\Sigma)^{-1} d\mbf{u}}\simeq 1-\frac{1}{2}d\mbf{u}^T \Sigma^{-1} d\mbf{u}
    \label{ec:app_factor2}
\end{equation}

Combining Eqs. (\ref{ec:app_factor1}) and (\ref{ec:app_factor2}) we get Eq. (\ref{ec:dif_fide_gaussian}).

\section{Quantum speed for pure states} \label{app:speed_pure}

In order to derive Eq. (\ref{ec:speed_unipure}) in the main text, we note that a general covariance matrix $\Sigma$ can be decomposed according to Williamson's theorem \cite{adesso2014} as $\Sigma = \hbar S K S^T$, where $S\in \mathrm{Sp}(2n)$ and \begin{equation}
    K = \bigoplus_{k=1}^n \left(\begin{array}{cc}
         \nu_k & 0  \\
         0 & \nu_k 
    \end{array}\right).
\end{equation}

The $\left\{\nu_i\right\}$ are the symplectic eigenvalues of $\Sigma$ such that the purity of the state is given by $\prod\limits_k \nu_k$. A state is pure if and only if all $\nu_k=1$. In that case, $\Sigma=\hbar S S^T$ and thus
\begin{equation}
    \Sigma^{-1}=\frac{1}{\hbar}\left(S^T\right)^{-1} S^{-1} \Rightarrow \Omega \Sigma^{-1} \Omega =-\frac{1}{\hbar}\Omega \left(S^T\right)^{-1}\Omega \Omega S^{-1} \Omega
    \label{ec:app_williamson}
\end{equation}
\noindent where we have exploited the fact that $\Omega^2=-\mathbb{I}_{2n\times 2n}$. Now, recall that by definition, a symplectic matrix $T$ is such that $T\Omega T^T=\Omega$. This condition can be rewritten as $\Omega T^{-1} \Omega=-T^T$. By evaluating $T=S$ and $T=S^T$, we obtain then that $\Omega S^{-1} \Omega = -S^T$ and $\Omega\left(S^T\right)^{-1}\Omega=-S$. Then, the Eq. (\ref{ec:app_williamson}) reads
\begin{equation}
    \Omega \Sigma^{-1} \Omega = -\frac{1}{\hbar} S S^T = -\frac{1}{\hbar^2}\Sigma,
\end{equation}

\noindent which is used to derive Eq. (\ref{ec:speed_unipure}) in the main text. \\

The Anandan-Aharonov relation \cite{anandan1990} states that the speed of unitary evolution for pure states is given by $\Delta E^2/\hbar^2$, where $\Delta E^2 \equiv \langle \hat{H}^2\rangle - \langle \hat{H}\rangle^2$. Thus, the expression for the speed in Eq. (\ref{ec:speed_unipure}) has to be equal to this quantity. Here we check that this is indeed true by direct calculation. First, recall that $\hat{H}=\frac{1}{2}\sum\limits_{ij} G_{ij} \hat{z}_i \hat{z}_j$, and so
\begin{equation}
    \hat{H} - \langle \hat{H}\rangle  = \frac{1}{2}\sum\limits_{ij} G_{ij}\left(\hat{z}_i \hat{z}_j - \langle \hat{z}_i \hat{z}_j \rangle\right).
\end{equation}
We can rewrite this expression in terms of the displaced quadrature operators $\delta \hat{\mathbf{z}}=\hat{\mathbf{z}}-\mathbf{u}$,
\begin{equation}
    \hat{H}-\langle \hat{H}\rangle = \frac{1}{2}\sum\limits_{ij}\left(\delta \hat{z}_i \delta \hat{z}_j - \langle \delta \hat{z}_i \delta \hat{z}_j \rangle + u_i \delta \hat{z}_j + u_j \delta\hat{z}_i\right)
\end{equation}
Then, we work out $\langle ( \hat{H} - \langle \hat{H}\rangle )^2\rangle$, and use the fact that expectation values of even powers of the $\delta \hat{z}_k$'s are zero, due to the Gaussian character of the state. The resulting expression reads
\begin{equation}
\begin{split}
   \langle ( \hat{H} - \langle \hat{H}\rangle )^2\rangle = \frac{1}{4} \sum\limits_{ijkl} G_{ij}G_{kl} \left( \langle \delta\hat{z}_i  \delta\hat{z}_j  \delta\hat{z}_k \delta\hat{z}_l\rangle - \langle \delta\hat{z}_i  \delta\hat{z}_j\rangle \langle \delta\hat{z}_k \delta\hat{z}_l\rangle + u_i u_k \langle \delta\hat{z}_j \delta\hat{z}_l\rangle \right. \\
   \left. + u_i u_l \langle \delta\hat{z}_j \delta\hat{z}_+\rangle + u_j u_k \langle \delta\hat{z}_i \delta\hat{z}_l\rangle + u_j u_l \langle \delta\hat{z}_i \delta\hat{z}_k\rangle  \right)
\end{split} 
   \label{ec:app_h2}
\end{equation}

The next step is to use Wick's theorem in order to write the fourth order moment in terms of the second order moments,
\begin{equation}
    \langle \delta\hat{z}_i  \delta\hat{z}_j  \delta\hat{z}_k \delta\hat{z}_l\rangle = \langle \delta\hat{z}_i  \delta\hat{z}_j  \rangle\langle\delta\hat{z}_k \delta\hat{z}_l\rangle + \langle \delta\hat{z}_i  \delta\hat{z}_k  \rangle\langle\delta\hat{z}_j \delta\hat{z}_l\rangle + \langle \delta\hat{z}_i  \delta\hat{z}_l  \rangle\langle\delta\hat{z}_j \delta\hat{z}_k\rangle,
\label{ec:app_wicks}
\end{equation}
\noindent and to note that 
\begin{equation}
    \delta \hat{z}_i \delta \hat{z}_j = \frac{1}{2}\left( \aconm{\delta \hat{z}_i}{\delta \hat{z}_j} + \conm{\delta \hat{z}_i}{\delta \hat{z}_j}\right)\rightarrow\langle \delta \hat{z}_i \delta \hat{z}_j\rangle = \frac{1}{2}\left(\Sigma_{ij}+i\hbar \Omega_{ij}\right)\equiv \frac{1}{2} V_{ij}.
\label{ec:app_V}
\end{equation}

With this elements in place, we now can combine Eq. (\ref{ec:app_h2}) with Eqs. (\ref{ec:app_wicks}) and (\ref{ec:app_V}). The resulting expression reads
\begin{eqnarray}
    \Delta E^2 &=& \frac{1}{16}\sum\limits_{ijkl} G_{ij}G_{kl}\left(V_{ik}V_{jl}+V_{il}V_{jk}\right)+\frac{1}{8}\sum\limits_{ijkl} G_{ij}G_{kl}\left(u_i u_k V_{jl} + u_i u_l V_{jk} 
     + u_j u_k V_{il}+ u_j u_l  V_{ik}\right) \\ &=& \frac{1}{8}\trace{GVGV^T}+\frac{1}{2}\mathbf{u}^T GVG\mathbf{u}=\frac{1}{8}\left(\traceb{(G\Sigma)^2} + \hbar^2 \traceb{(G\Omega)^2}\right) + \frac{1}{2}\mathbf{u}^T G\Sigma G\mathbf{u},
\end{eqnarray}

\noindent where we have used that $\mathbf{u}^T G\Omega G\mathbf{u} = 0\: \forall \mathbf{u}$ since $G\Omega G$ is an antisymmetric matrix.

\section{QSL for single mode Gaussian unitary evolution} \label{app:singlemode}

For $n=1$, the most general covariance matrix can be written as
\begin{equation}
    \Sigma=\hbar c\, R(\theta)DR(\theta)^T,\ \mathrm{with}\  R(\theta)=\left( \begin{array}{cc}
         \cos \theta & \sin \theta  \\
         -\sin \theta & \cos \theta 
    \end{array}\right)
    \label{app_ec:general_state}
\end{equation}
and $D=\mathrm{diag}(\ex{r},\ex{-r})$. For this analysis we will focus on the role of squeezing, and thus we will consider undisplaced states ($\mathbf{u}=0$). The Hamiltonian generating the evolution is an element of the algebra $\mathfrak{sp}(2)$, which has dimension $2n-1=3$ and its spanned by the elements
\begin{equation}
    G_0=\left( \begin{array}{cc}
         1 & 0  \\
         0 & 1 
    \end{array}\right),\ G_1=\left( \begin{array}{cc}
         1 & 0  \\
         0 & -1 
    \end{array}\right),\ G_2=\left( \begin{array}{cc}
         0 & 1  \\
         1 & 0 
    \end{array}\right).
\end{equation}
So, we consider dynamics driven by the most general generator
\begin{equation}
    G = g_0 G_0 + g_1 G1 + g_2 G_2 \equiv g_0 G_0 + g_S G_S(\phi) 
    \label{app_ec:general_hami}
\end{equation}
where we introduced the alternative parametrization $G_S(\phi)=R(\phi)G_2 R(\phi)^T$,  $g_1=g_s\sin(2\phi)$ and $g_2 = g_s\cos(2\phi)$. In this notation, $g_0$ is the weight of the number-preserving part of the Hamiltonian, while $g_S$ is the weight of the number-non-preserving part of $G$. Using the expressions in Eqs. (\ref{app_ec:general_state}) and (\ref{app_ec:general_hami}), we can evaluate the speed in Eq. (\ref{ec:speed_unitary}). After some algebraic manipulation, the result reads
\begin{equation}
V_\mathrm{U}^2=\frac{1}{2}\left[\left(g_0 \sinh(r)-g_S\sin(2\delta)\cosh(r)\right)^2+g_S^2 \cos^2(2\delta) \right]
\label{app_ec:vel_singlemode}
\end{equation}
\noindent where we introduced $\delta=\theta-\phi$.\\

Our goal is to derive, for a given value of squeezing, the Hamiltonian that maximizes and minimizes the speed. Differentiating Eq. (\ref{app_ec:vel_singlemode}) with respect to $\delta$ and equating it to zero reveals the existence of the following extrema in the interval $\delta\in [0,\pi]$:
\begin{equation}
    \delta_c^{(1)}=\frac{\pi}{4},\ \mathrm{and}\  \delta_c^{(2)}=\frac{3\pi}{4}
\end{equation}

\noindent for all values of parameters, while two extra extrema $\delta_c^{(\pm)}$ appear if $\frac{g_0}{g_S} \leq \tanh(r)$, which obey the equation
\begin{equation}
    \sin\left(2\delta_c^{(\pm)}\right)=\frac{g_0}{g_S}\coth(r)
\end{equation}

Straightforward stability analysis reveals that $\delta_c^{(2)}$ is always a maximum (and furthermore, its global), $\delta_c^{(\pm)}$ are always minima (when they exist), and thus $\delta_c^{(1)}$ is a minima for $g_0/g_S> \tanh(r)$ and a maxima otherwise.  The resulting situation is depicted in Fig. \ref{fig:singlemode}. 

Let us now discuss the maximum speed, which is given by
\begin{equation}
    V^2_{\mathrm{max}}=V^2_\mathrm{U}\left(\delta_c^{(2)}\right)=\frac{1}{2}\left(g_0\sinh(r)+g_S\cosh(r)\right)^2
\end{equation}
For a fixed degree of squeezing $r$, the above expression reaches its maximum value $g^2\, \cosh(2r)/2$ when $g_0/g_S=\tanh(r)$. Here $g$ is taken to be some overall Hamiltonian strength which we assume to be fixed, i.e. $g_0=g\:\tilde{g_0}$ and $g_S=g\:\tilde{g_s}$. This result tells us that we can always achieve a maximum speed proportional to $\cosh(2r)\sim \ex{2r}$ by using an optimal choice Hamiltonian. If $r$ is small, the choice is to set $g_S \gg g_0$. For large $r$, on the other hand, the best choice is to set $g_S\simeq g_0$. 

Conversely, one can analyze the minimum possible speed $V^2_{\mathrm{min}}$ for these systems. In this case, the expression for $V^2_{\mathrm{min}}$ is different depending on the relation between $g_0$ and $g_S$. If $g_0>g_S \tanh(r)$, then the minimum occurs at $\delta_c^{(1)}$ and we obtain
\begin{equation}
    V^2_{\mathrm{min}}=V^2_\mathrm{U}\left(\delta_c^{(1)}\right)=\frac{1}{2}\left(g_0\sinh(r)-g_S\cosh(r)\right)^2
\end{equation}
This is the naturally opposite situation as the one described before. The minimum possible speed is 0, and its achieved for $g_S/g_0=\tanh(r)$; for low squeezing, the optimal choice is one where $g_0\gg g_S$. At higher squeezing, the optimal choice is $g_0\simeq g_S$ as before. If $g_0 < g_S \tanh(r)$, then the minimum occurs at both $\delta_c^{(\pm)}$, for which
\begin{equation}
    V^2_{\mathrm{min}}=V^2_\mathrm{U}\left(\delta_c^{(\pm)}\right)=\frac{1}{2}\left(g_S^2-g_0^2\right)
\end{equation}
Interestingly, even when the speed cannot be turned to zero in this parameter regime, it will always be independent of the amount of squeezing in the system.

\section{Quantum speed for open system dynamics} \label{app:open_quantum}

Markovian Gaussian-preserving evolution can be shown to lead to the following general equations of motion for a $n$-mode system \cite{genoni2016}
\begin{eqnarray}
\dot{\Sigma} &=& B\Sigma + \Sigma B^T + D \\
\dot{\mbf{u}} &=& B\mbf{u}
\end{eqnarray}

Here $B$ and $D$ are generic $2n\times 2n$ matrices which obey the relation $D+i\Omega B_a \Omega ^T \geq 0$ where $B_a =\Omega^T B-B^T\Omega$. Its convenient to write $B=\Omega(G+F)$, where $G$ is symmetric (i.e. corresponding to the unitary dynamics, $\Omega G = A$) and $F$ is antisymmetric. For a single mode system, $n=1$, expressions simplify considerably since the most general antisymmetric matrix can be written as
\begin{equation}
    F=\left(\begin{array}{c c}
    0 & g/2 \\
    -g/2 & 0 \end{array}\right)
\end{equation}

Then, the generator takes the form $B=\Omega(G+F)=A + \Omega F =  A-\frac{g}{2}\mathbb{I}$. By defining $M=D/g$, this proves Eq. (\ref{ec:open_eqs}) in the main text. Notice that the condition over $D$ reads $\det D\geq \det (2F)=g^2$ and so $\det M \geq 1$.

The quantum speed arising from the evolution of the covariance matrix is
\begin{equation}
    V^2_{\mathrm{cov}}=\frac{1}{16} \tr{\left(\Sigma^{-1} \,\dot{\Sigma}\right)^2}
\end{equation}

Since $\dot{\Sigma}$ is the sum of a unitary and a nonunitary contribution, and given the quadratic dependence of the speed, naturally one obtains $V^2_{\rm cov}\!=\!V^2_{\rm U} + \chi_{\rm NU}$, i.e. a contribution solely from unitary dynamics plus a nonunitary correction, which itself can be thought of as a combination of purely nonunitary and a cross term.

\end{document}